# Nanogaps by direct lithography for high-resolution imaging and electronic characterization of nanostructures

Michael D. Fischbein[a] and Marija Drndić

*Department of Physics and Astronomy, University of Pennsylvania, 209 South 33rd Street,*

*Philadelphia, PA 19104*

We report a method for fabricating nanogaps directly with electron beam lithography (EBL). The primary resolution-limit of EBL, electron back-scattering, is reduced dramatically by using a thin-film as a substrate. We show that this resolution enhancement allows one to fabricate metal electrodes with separation from arbitrarily large to under one nanometer. Furthermore, because these nanogaps are on a thin film, they can be imaged with high-resolution transmission electron microscopy (HRTEM). Using these nanogaps we measured the charge transport through several coupled PbSe nanocrystals and correlated the data with detailed structural information obtained by performing HRTEM on the same device.

[a]Electronic Mail: mlfisch3@physics.upenn.edu

Efforts toward achieving electrical contact to nanostructures have been active for over a decade.[1] Even though several devices based on "nanogaps" – two wires separated by a nanometer-scale distance - have been demonstrated,[2] their realization has remained a significant challenge. Even the best methods are highly labor intensive and suffer from low yield and poor geometrical control. Most nanogaps are also incompatible with high-resolution transmission electron microscopy (HRTEM). As a consequence, the proof of the nanogaps' quality and content in past studies has been indirect. Moreover, interesting quantum effects, such as Coulomb blockade and Kondo effects, have now been reported in electromigrated-breakjunction gaps containing *no sample* – only metallic debris produced from the fabrication.[3] High-resolution imaging is therefore required to ensure the quality of nanogaps and to be able to identify possible artifacts.

In this letter, we report the successful fabrication of sub-nanometer size gaps on thin membrane substrates directly with electron beam lithography (EBL). Because these nanogaps are made on thin films, it is easy to examine their structure and content with HRTEM. We have applied these nanogaps to the investigation of systems of very few coupled PbSe nanocrystals and results are discussed below.

To fabricate membrane substrates we used a recipe in which doped silicon ($Si^+$) wafers with highly polished, 100 nm-thick low-stress silicon nitride ($Si_3N_4$) layers grown on both sides are first coated with photoresist. Photolithography is then used to remove a square region of the resist, thereby revealing the $Si_3N_4$ underneath. This side of the wafer is then exposed to a $SF_6$ plasma etch, which removes a square of $Si_3N_4$, revealing the $Si^+$ underneath. Finally, the wafer is exposed to a KOH wet etch. The KOH etches anisotropically through the $Si^+$ along lattices planes until the $Si_3N_4$ on the other side of the wafer is revealed. A free standing membrane window, in our case ~ $(50 \ \mu m)^2$, of 100nm thick $Si_3N_4$ is therefore defined.

We performed the EBL on these thin $Si_3N_4$ membrane substrates with a thermal-emission JEOL 6400 scanning electron microscope (SEM) operating at its maximum accelerating voltage of 30 kV. The electron beam (smallest attainable diameter ~ 30 nm) was controlled with a Raith writing program. A 1 $\mu m$ thick layer of C2 PMMA (950 molecular weight PMMA, 2% in chlorobenzene) was spin-cast onto the $Si_3N_4$ windows at ~5000 rpm to achieve a resist layer of ~ 100 nm. Nanogaps were written using a 10 pA electron beam in the EBL chamber with pressure below $10^{-6}$ Torr. The resist was exposed at a magnification of 2000x, corresponding to a write field of $(40 \ \mu m)^2$, to beam doses ranging from 1000 to 2000 $\mu C/cm^2$, depending on the desired size and geometry of the nanogap. Larger features were then written into the resist with standard EBL parameters for the purpose of later connecting the nanogaps to large contact pads. The device was then developed in MIBK:IPA (3:1 volume ratio) for 60 seconds and loaded into the low pressure environment of a thermal evaporator. For metallization, several nanometers of either nickel or chromium were evaporated first to act as an adhesion layer between the gold and the substrate, followed by 30 nm of gold. The wafers were put into acetone at room temperature in order to achieve lift-off. The gaps were then imaged with JEOL 2010 and JEOL 2010F Transmission Electron Microscopes.

Figure 1 shows TEM images of examples of nanogaps on $Si_3N_4$ membranes with gap sizes 0.7, 1.5, 3, 4, 5 and 6 nm (Fig 1 (a-f), respectively). A HRTEM image of another 4 nm gap is also shown (Fig. 1 (g)). Our nanogap fabrication process is high-yielding and we have controllably made hundreds of nanogaps from as small as < 1 nm to arbitrarily large size with step sizes < 1 nm (images of additional nanogaps are available by request). These nanogaps are robust at room temperature and do not change over time.

We explain this successful fabrication as the result of minimizing electron back-scattering during the EBL processing. Electron back-scattering is the primary factor limiting the feature resolution of EBL. Efforts to understand and reduce electron back-scattering in EBL date from the 70s and 80s, with most contributions made by IBM.[4] Interestingly, despite these efforts and the acquired fundamental knowledge, the potential of this knowledge has not been explored for realizing nanogaps. Moreover, recent efforts of making nanogaps compatible with TEM characterizations surprisingly resorted to breakjunction techniques without first exploring the limitations of standard EBL.[5]

Standard EBL is a several step process. In short, a layer of "electron-resist" (PMMA) is spin-cast onto a wafer and exposed to an electron beam in targeted areas. The energetic electrons break the PMMA bonds and this soluble resist is dissolved. The remaining resist acts as a stencil of the lithographically defined pattern. Devices are made by metal deposition and a "lift off" of the undesired metal by dissolving the resist underneath it. The final device is composed of metal features defined by the EBL pattern.

Figure 2 (a) is a schematic of the electrons' trajectories in a thick resist-insulator-silicon substrate. The incident electron beam passes through the PMMA resist and the insulator (in this example, $SiO_2$) without affecting most of the resist bonds. When these "forward-scattered" electrons reach the doped silicon, strong scattering with the silicon lattice cause many to reflect backward with a wide spatial distribution and energies similar to that of the incident beam. The size of this burst of back-scattered electrons largely determines the size of the feature written. Worse still, areas intended to remain unexposed can get a large dose of these energetic electrons. This phenomenon is known as the "proximity effect".[6]

If the distributions of scattered electrons (Fig. 2 (a)) are approximated as Gaussian, with "characteristic widths", $\sigma_f$ and $\sigma_b$, for the forward-scattered and back-scattered electrons, respectively, then the distribution of resist exposure is given by the energy deposition function (EDF),[7] $F(r) = k\left\{\exp\left[-\left(\frac{r}{\sigma_f}\right)^2\right] + \eta\left(\frac{\sigma_f}{\sigma_b}\right)^2 \exp\left[-\left(\frac{r}{\sigma_b}\right)^2\right]\right\}$.

Here, $r$ is the radial distance from the exposure center, $\eta$ is the ratio between the energy dissipation of back-scattered and incident electrons along the beam axis and $k$ is a constant used to normalize $F(r)$ to 1 at $r = 0$. For a 500 nm-thick layer of PMMA resist on doped silicon and using a 25 keV incident beam, $\sigma_f$, $\sigma_b$ and $\eta$ are 0.06 µm, 2.6 µm and 0.51 respectively.[7]

Figure 2 (b) is a schematic of a thin $Si_3N_4$ membrane used in this work. Here, the proximity effect is reduced dramatically, $\sigma_b \sim 0$, and the EBL feature resolution is limited by the much smaller value $\sigma_f$ and the EDF becomes $F(r) \sim \exp\left[-\left(\frac{r}{\sigma_f}\right)^2\right]$.

Conveniently, the distribution of forward-scattered electrons decreases with increasing $r$ even *faster* than a Gaussian and is better approximated by multiple small-angle scattering events.[8] Because of the extremely narrow distribution of electrons in the absence of back-scattering, it is possible to expose two nearby regions while leaving a nanometer-scale gap between them unexposed (Fig. 2 (b)). Our results thus demonstrate that electron back-scattering can be sufficiently minimized to make nanogaps efficiently and down to < 1 nm. Our results are not necessarily limited to thin substrates because electron back-scattering can, in principle, be sufficiently reduced on appropriately processed thick substrates. However, as we show here, the ability to image nanogaps and their content with HRTEM is a valuable tool for the fundamental characterization of nanostructures and this capability is only possible with thin substrates.

Figure 3 (a) is a TEM image of several 6.4 nm diameter PbSe nanocrystals (NCs) capped with oleic acid localized in an 11 nm gap. PbSe NCs behave as quantum dots (QDs) and display quantum confinement even at room temperature. Coupled QDs are interesting for probing charge correlation effects, as well as for applications, such as quantum computing.[9]

Figure 3 (b) shows the current-voltage (I-V) characteristic at T = 5 K corresponding to the *same* nanogap shown in Fig. 3 (a), before and after PbSe NC deposition. The current measured after imaging the bare gap and before adding NCs (red curve) is immeasurably small (< 30 fA). In contrast, the I-V curve for the added PbSe NCs (black curve) shows a rich structure that was reproducible for each voltage sweep. In particular, there is a clear voltage threshold, $V_{th} \approx 53$ mV, for the transition from zero to finite current at positive bias. Figure 3 (c) is a HRTEM image which focuses on the upper region of the gap area. It shows with unprecedented clarity the detailed lattice structure of 4 closely packed PbSe NCs between two Au electrodes. We note that only two NCs are actually touching the contacts. From several HRTEM images focusing on different parts of this device, we determined this area to be of primary interest. No other region of the nanogap contained NCs that completely bridged the two electrodes and were also close enough to each other for electron tunneling to occur between them. The inter-NC charging energy is approximated to be $E_c \approx \frac{e^2}{2C}$, where for a given NC radius, $r$, and inter-NC spacing, $2d$, the capacitance between neighboring NCs separated by a medium with dielectric constant, $\varepsilon$, is $C \sim 2\pi\varepsilon_0 \varepsilon r \ln\left(\frac{r+d}{d}\right)$. From Fig. 3 (c) we obtained the NC diameter (6.4 nm) and the inter-NC spacings between nearest-neighbors for each NC (from 1 to 2.3 nm). Using the dielectric constant of oleic acid (~2), this gives a value $E_c = 56\ meV$, which agrees very closely with the measured value for $V_{th}$.

In conclusion, we have discovered an efficient route to fabricating high-quality nanometer-size electrode gaps compatible with atomic resolution imaging. The fabrication of these nanogaps is direct and can be achieved by anyone with access to a standard EBL system. The importance of nanogaps is very broad. They promise to have applications in many fields: from evolutionary biology to quantum computing. In addition to their usefulness, nanogaps represent a landmark in the advancement of modern science towards bridging the classical and quantum worlds.

This work was supported by the ONR Young Investigator Award (N000140410489), the American Chemical Society (ACS) PRF award (41256-G10), and in part, the NSF Career Award DMR-0449553 and NSF MRSEC DMR00-79909. M.F. acknowledges funding from the NSF-IGERT program (Grant DGE-022166).

.

FIG. 1

TEM image of nanogaps with sizes 0.7 nm (a), 1.5 nm (b), 3 nm (c), 4 nm (d), 5 nm (e) and 6 nm (f). (g) HRTEM image of another 4 nm nanogap. The crystal lattice planes of the electrodes (g) are seen clearly. Inset to (a): SEM image of a full device consisting of electrodes (white lines) on a suspended 50 μm x 50 μm $Si_3N_4$ membrane (black square) and connected to larger wires. Inset to (c): TEM image of electrodes (black lines) on a suspended $Si_3N_4$ membrane.

FIG. 2

(a) Standard PMMA-$SiO_2$-$Si^+$ substrate. An incident electron beam "forward-scatters" slightly in the PMMA and $SiO_2$ layers. Strong scattering in the $Si^+$ results in broadly distributed "back-scattered" electrons which expose a wide region of the PMMA. (b) PMMA-$Si_3N_4$ substrate used to make nanogaps with EBL. Two nearby areas are shown being sequentially exposed to an electron beam while the small "nanogap" region between them is left unexposed.

FIG. 3

(a) TEM image of a small array of PbSe NCs inside of an 11 nm nanogap defined by two Au electrodes. (b) I-V characteristic at T = 5 K in high vacuum for the same device in a without NCs (red curve) and with NCs (black curve). The threshold voltage for positive forward bias is marked at 53 mV. A TEM image of the device before depositing NCs is also shown (inset, scale bar is 20 nm). (c) HRTEM image of the upper portion of the gap showing the detailed lattice structure of four PbSe NCs that dominate the charge transport across the gap.

Figure 1

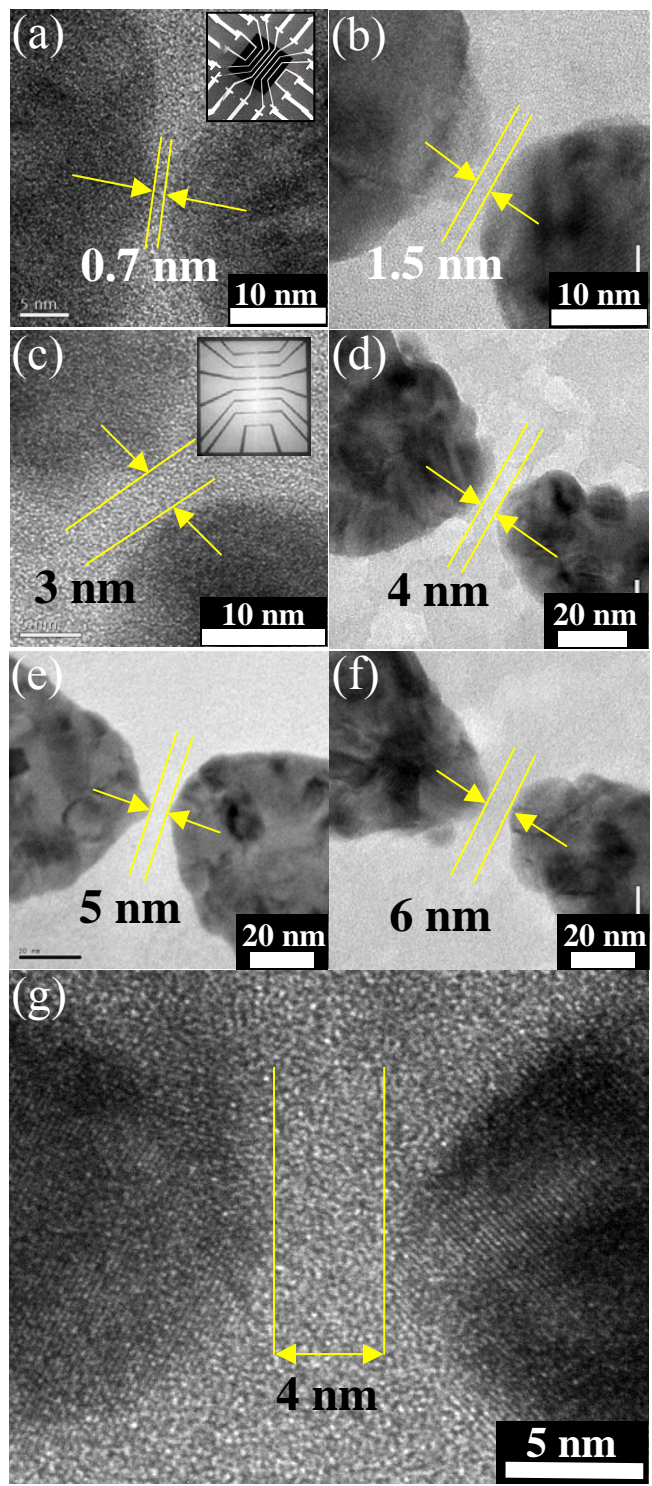

Figure 2

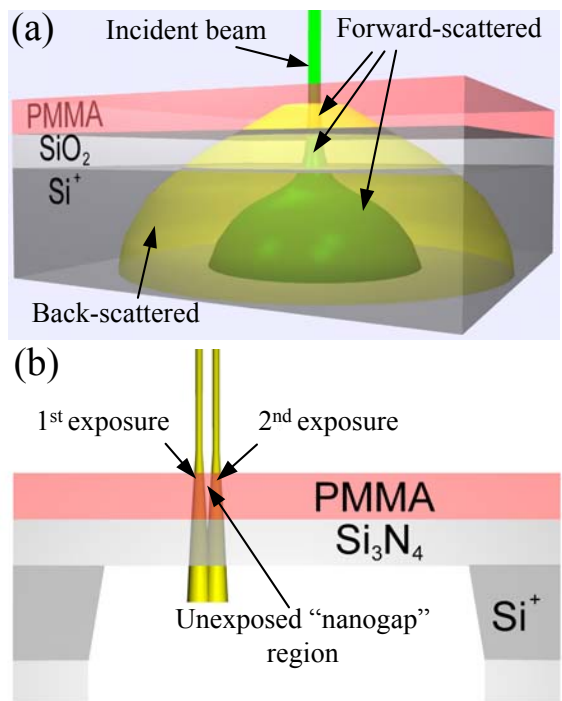

Figure 3

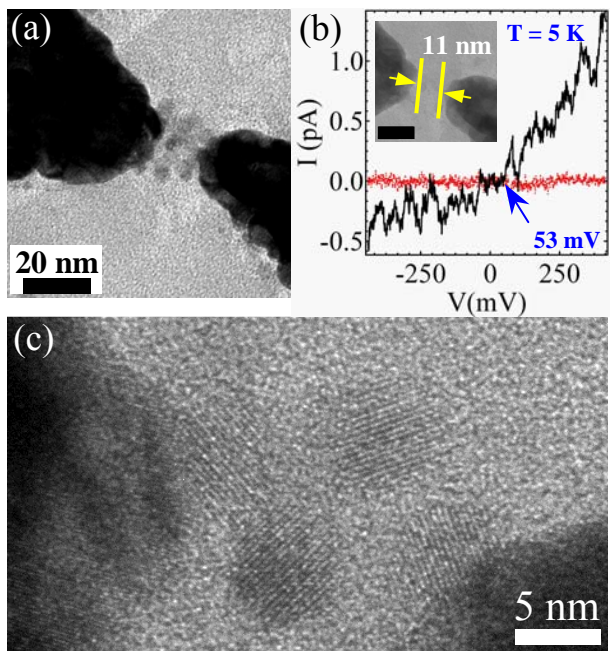